# David Bohm, Roger Penrose,

# and the Search for Non-local Causality

Before they met, David Bohm and Roger Penrose each puzzled over the paradox of the arrow of time. After they met, the case for projective physical space became clearer.

*** 

A machine makes pairs (I like to think of them as shoes); one of the pair goes into storage

before anyone can look at it, and the other is sent down a long, long hallway. At the end

of the hallway, a physicist examines the shoe.[1] If the physicist finds a left hiking boot, he

expects that a right hiking boot must have been placed in the storage bin previously, and

upon later examination he finds that to be the case. So far, so good. The problem begins

when it is discovered that the machine can make three types of shoes: hiking boots, tennis

sneakers, and women's pumps. Now the physicist at the end of the long hallway can

randomly choose one of three templates, a metal sheet with a hole in it shaped like one of

the three types of shoes. The physicist rolls dice to determine which of the templates to

hold up; he rolls the dice after both shoes are out of the machine, one in storage and the

other having started down the long hallway. Amazingly, two out of three times, the

random choice is correct, and the shoe passes through the hole in the template. There can then be a test of the pairing. Even more amazing is that 100 percent of the time that there is such a test, the shoe placed earlier in the bin is the pair to the shoe caught—either a left or right boot, a left or right tennis sneaker, a left or right pump, as the case may be. The shoe in the bin "knew" how the delayed choice would turn out. Since the shoe placed earlier in the bin is the result of the choice made later, the effect precedes the cause.

---

[1] This rendition of the Einstein-Podolsky-Rosen, 1935, delayed-choice thought experiment is paraphrased from Bernard d'Espagnat, 1981, with modifications via P. K. Aravind, 1997.  See Bernard d'Espagnat, "The Concepts of Influences and of Attributes as Seen in Connection with Bell's Theorem, *Foundations of Physics* 11 (1981): 205–34; P. K. Aravind, "Borromean Entanglement of the GHZ State." in *Quantum Potentiality, Entanglement, and Passion-at-a-Distance*, ed R.S. Cohn, M. Horne, and J. Satchel (Dordrecht: Kluwer, 1997. I want to thank Dr. Aravind for many long discussions and follow-up email over a dozen years.

During the 1940s and 1950s, it was assumed that such a delayed-choice experiment would eventually be built, and that if the physics of quantum mechanics were correct and complete, such a paradoxical result would be manifest. The violation of causality troubled physicist David Bohm. Without a clear definition of cause and effect, there is no time ordering of events. Without a clear arrow of time, there is no entropy, and without that, there is no physics at all.

It is as if you filled the car with gas, drove around all day, and then, when you tried to put more gas in at the end of the day, found the tank was full, even topped up. The potential energy of the gas was converted to the kinetic energy of the car's motion, and because there is at least some loss due to friction in the machinery, the conversion is one way; an arrow of time is defined. It seems that the time order of events, the arrow of time, is unavoidably bound up in the most basic principle of physics.

The Copenhagen interpretation of quantum physics has a solution to the paradox: there is no problem with the time order of events because the shoe (or particle) that is at the end of the hallway is observed first, and only then is the shoe in storage observed. Nothing is really real until it is observed—a statement much easier to accept when one considers particles as extended waves, rather than as something concrete and located precisely in space, like a shoe. Yet, in some sense, the Copenhagen interpretation requires the individual human mind to make the world, and it did not satisfy Bohm. He felt that the theory was incomplete. There must be a hidden variable, a common denominator that tied these two real and separate particles together, and thus explained their coordinated behavior.

David Bohm was born in 1917 into an immigrant Jewish family in the hardscrabble mining town of Wilkes-Barre, Pennsylvania, where he developed sympathy for the working-class families who visited his father's furniture store. In 1943 he received his Ph.D. from Berkeley, working under J. Robert Oppenheimer. Known for left-wing views (he was a member of the Communist Party from 1942 to 1943), Bohm was denied security clearance to work at Los Alamos on the atomic bomb project, even though his papers were of considerable value to that effort. In 1949 Bohm's name came before the House Committee on Un-American Activities, and he refused to cooperate. He was eventually cleared of contempt by the Supreme Court, but in the McCarthy era, the taint was enough to deny him reappointment at Princeton University, where he had taught successfully since 1947 and had been admired by Albert Einstein, and where his 1951 book *Quantum Theory* had become an influential textbook. Nevertheless, Harold W. Dodds, the president of Princeton and an ardent anti-Communist, vigorously fought Bohm's reappointment, confusing the man's politics with his physics.

Although the facts are not in dispute, historians disagree on the effect that Princeton's rejection had on Bohm. He left America, went to Brazil, gave up his American citizenship, lived briefly in Israel, and then went to Birkbeck College at the University of London. In Brazil (where he developed stomach trouble that plagued him all his life), Bohm was dissatisfied with the technical capabilities available to him, and his work was never again considered mainstream. Science historian Russell Olwell paints a picture of Bohm as unfairly persecuted and professionally marginalized, a loss to the field of physics. Brazilian historian Olival Freire Jr., however, paints a different picture. Bohm had Brazilian graduate students at Princeton and through their intervention was set up quite nicely in Säo Paolo. He had budgets to hold international conferences that were attended by Richard Feynman and Isidor Rabi, among other important physicists. Giving up his American citizenship was merely a technical requirement for obtaining a Brazilian passport in order to work in Israel, where he had an important appointment at the Technion and could work with better tools. Asked but not answered in Freire's paper, however, is the question as to whether or not Bohm's stubborn insistence on a hidden variable (he was almost alone in this belief) was somehow influenced by the way he was treated by the American authorities.

In his 1957 book *Causality and Chance in Modern Physics*, Bohm reviewed the well-known Young double-slit experiment of 1801. If two slots are open, a beam of electrons (light in the original experiment) makes an interference pattern as though the beam were a wave. If only one slit is open, the beam of electrons still makes a distribution pattern as though it were a wave, even though it is made up of discrete particles. (Bohm stated [pp. 114–15] that this is caused by random fluctuations in the system.) He wrote that there must be something that is neither particle nor wave that permits both: "an interesting and suggestive possibility is then that of a sub-quantum mechanical level containing hidden variables" (p. 101). Such hidden variables would bring back a more classic concept of causality. "It would appear, therefore, that the conclusions concerning the need to give up the concepts of causality, continuity of motion, and the objective reality of individual micro-objects have been too hasty" (p. 95). Such hidden variables could be present yet are impossible to directly detect—by definition being below the scale of Heisenberg indeterminacy. "For even if a sub-quantum level containing 'hidden' variables of the type described previously should exist, these variables would then never play any real role in the prediction of any possible kind of experimental results" (p. 85).[2]

In Israel in 1957, Bohm and his student Yakir Aharonov published a new non-local effect, commonly called the A-B effect. In a region of space isolated from electric and magnetic fields (inside a solenoid, for example), a charged particle could still feel the effects of a force from outside the barrier. Such a force was previously described only as a mathematic potential, yet it was now observed to have a physical effect at a distance. As P. K. Aravind put it to me in an email on December 4, 2013:

> Now there are regions where the magnetic field vanishes but the vector potential that gives rise to it does not. A and B showed that a particle (such as an electron) that can move in the region where the magnetic field vanishes but the vector potential does not can be affected by the vector potential in a way that can be measured by experiment. This shows that the vector potential has a physical significance, which went counter to the conventional wisdom at that time.

Bohm continued to probe the concept of non-locality and discovered a way to test EPR directly, by measuring the spins of the entangled pairs along one of three axes, chosen at random.

---

[2] Perhaps in sympathy with his subject matter, Bohm's argument flows back and forth in his text, frequently doubling back on itself; note the order of these quotes.

In 1964 the Irish physicist John Bell, reading Bohm, devised a way to test EPR as modified by Bohm. But Bell went further: there would be a slightly different result if Bohm's hidden variables were present. As the Bell inequality experiments were repeated by different teams, including work done by others at Birkbeck College, as the "hallway" gets longer and the time interval between the two observations of the "shoes" gets shorter, Bohm was forced to conclude that the entangled particles are both real and non-local and that there is no hidden connection between the pairs. But he still could not let matters lie; he could not give up causality altogether. So he began a new definition of space, what he called a new order in physics that challenges the very idea of time and location. It is hard to summarize all that Bohm meant by the word *order*—a new paradigm we might say, but more than that, a new specific model.

Bohm wanted to dispense with the problem of the time order of events by demoting the entire concept of time to something of a superficial construct (in this he anticipated some contemporary thought on the matter): "if an order is 'enfolded' throughout all of space and time, it cannot coherently be regarded as constituting a signal that would propagate information from one place to another over a period of time."[3] The same demotion to "high-level abstractions" is given to the other nuggets of physics: "we do not regard terms like 'particle,' 'charge,' 'mass,' 'position,' 'momentum,' etc., as having primary relevance in the algebraic language [that defines events]."[4]

Thus, the word "electron" should be regarded as no more than a name by which we call attention to a certain aspect of the holomovement, an aspect that can be discussed only by taking into account the entire experimental situation and that cannot be specified in terms of localized objects moving autonomously through space. And, of course, every kind of "particle" which in current physics is said to be a basic constituent of matter will have to be discussed in the same sort of terms (so that such "particles" are no longer considered as autonomous and separately existent). Thus, we come to a new general physical description in which "everything implicates everything" in an order of undivided wholeness.[5]

---

[3] David Bohm, "Quantum Theory as an Indication of a New Order in Physics. B. Implicate and Explicate Order in Physical Law," *Foundations of Physics* 3 (1973): 164.
[4] Ibid., p. 160.

It might seem that Bohm finally capitulated to the Copenhagen interpretation—that matter exists only as extended waves until it is localized by observation. Except that Bohm had a new model of space, the hologram. Whereas the Copenhagen interpretation is a discussion of the nature of matter, Bohm's discussion is about the nature of space: it is multiply-connected as though it were a hologram. All points in space contain information enfolded in it about all other points of space. Tear a hologram film of an apple into small pieces; unlike a photographic negative, each small piece of the hologram contains a picture of the whole apple. Bohm states that space is like this. David Peat, in an email to me dated October 13, 2013, is uncomfortable with such a claim: "In order to explain to non-physicists he developed a couple of examples, [one of which was the hologram] which he would always stress where not the implicate order themselves but pointed to it." Yet Bohm does not appear to be describing a mere metaphor.

When distinguishing the difference between a hologram and a lens, Bohm says about the hologram: "There is the germ of a new notion of order here. This order is not to be understood solely in terms of regular arrangements of *objects* (e.g., in rows) or as a regular arrangement of *events* (e.g., in a series). Rather, a *total order* is contained, in some *implicit* sense, in each region of space and time.[6]

[5] Ibid., p. 153.

<p style="text-align:center">***</p>

Meanwhile, Roger Penrose was working on another paradox of time order and entropy. As presented by David Layzer,[7] the problem can be understood by imagining a bottle of perfume open in a closed room. Little by little the perfume evaporates; the room is filled with the smell and no perfume remains in the bottle. The paradox is that if a very short film were made focusing on just a few molecules of perfume as they crash into one another and bounce around, a physicist could not tell if the film were running forward or backward; the molecules can bounce down just as easily as they bounce up. There is, then, a disconnect between the global picture, which shows entropy (the potential energy stored in the bottle of perfume is converted to the kinetic energy of the evaporation—a one-way conversion that constitutes an arrow of time), and the local picture in which actions are time-symmetric. It will likely be the case that a few molecules of perfume will bounce back into the bottle, but overall no one would wait for all of the perfume to be back in the bottle.

Roger Penrose was born into a distinguished family in 1931. His parents were doctors, his older brother a mathematician, his younger brother a chess champion, and the artist and Picasso scholar Roland Penrose is an uncle. Unlike Bohm's unhappy career, Penrose's went from honor to honor. He graduated from University College, London, with first class honors in mathematics, and he received a NATO Research Fellowship for study in the United States and, later, the Rouse Ball Professorship at Oxford. He published many important papers and technical books, as well as such popular books as *The Emperor's New Mind* (1989). Prizes, visiting professorships, invitations to conferences—Penrose has been lauded his entire professional life.

---

[6] Ibid., p. 147.
[7] David Layzer, "The Arrow of Time," *Scientific American* (1975): 56–69.

On a small scale, many events in physics are time-symmetric, and Penrose puzzled over the "splitting" and sought a natural way to mathematically model time reversibility. He found that natural way in an alternate formalism he discovered to model the Lorentz transformations, which describe the effects of special relativity. Penrose showed that special relativity transformations could be considered to be projective transformations on the Riemann Sphere (specifically *Möbius transformations*), a geometric model that represents all the complex numbers together with $\infty$. Around the equator of this sphere are all the real numbers from 0 to infinity, both the positive numbers to infinity and the negative numbers to infinity. A longitudinal cut through the poles shows all the positive and negative imaginary numbers, $i$ (the square root of minus one). If homogeneous coordinates are used to describe rays from the center of this sphere, then an arbitrary chunk of space-time, with its + + + - Minkowski signature, can be represented as a region of this Riemann Sphere. Here is the example Penrose often gives:

Imagine an observer situated at a point in spacetime, out in space looking at the stars. Suppose she plots the angular position of these stars on a sphere. Now, if a second observer were to pass through the same point at the same time, but with a velocity relative to the first observer then, owing to aberration effects, he would map the stars in different positions on the sphere. What is remarkable is that the different positions of the points on the sphere are related by a special transformation called a *Möbius transformation*. Such transformations form precisely the group that preserves the complex structure of the Riemann sphere. Thus, the space of light rays through a spacetime point is, in a natural sense, a Riemann sphere.  I find it very beautiful moreover, that the fundamental symmetry group of physics relating observers with different velocities, the (restricted) Lorentz group, can be realized as the automorphism group of the simplest one-(complex-) dimensional manifold, the Riemann sphere.[8]

 Projections from the north pole are considered positive, pointing toward the future. Projections from the south pole are considered negative, pointing to the past. Thus a simple geometric interpretation is given to the splitting of time direction in a sophisticated model of physical space-time.

　　　Consistently throughout his writings, Penrose has emphasized that light rays be thought of as projective elements. The very good physical reason for this is that a light ray has a direction but no fixed length: from its point of view, it is wherever it was and where it will ever be at the same instant. Of course, slower-moving observers see things differently. I see a photon that was created in a nuclear explosion on Alpha Centauri that traveled 4.37 years until it is absorbed, and destroyed, in my eye, but for the photon itself no time has passed, no distance was traveled. Light rays are projective points and make the space of special relativity a projective space.

　　　There are three ways mathematicians describe the projective point, and all three give partial insight to this rich concept: by drawing a circle where one ends up at the starting point; establishing homogeneous coordinates, which are given as ratios rather than as fixed numbers (the collapsing of all points together—what an artist would call the completely foreshortened line); and identifying the end points, that is gluing the front and back together. Such a gluing can induce a twist, making a vector into a spinor.

[8] Stephen Hawking and Roger Penrose, *The Nature of Space and Time* (Princeton, NJ: Princeton University Press, 1996), p 109. See also Roger Penrose and Wolfgang Rindler, *Spinors and Space-Time: Volume 1, Two-Spinor Calculus and Relativistic Fields* (Cambridge, UK: Cambridge University Press, 1984).

Consider the Poincaré dodecahedron. A dodecahedron is a platonic solid made up of twelve pentagons in six sets of parallel faces. It can be thought of as a six-axis antiprism because each face is rotated with respect to its parallel pair. The Poincaré dodecahedron glues (or identifies) each set of parallel faces, twisting the object in the process. There is serious speculation on the part of Jean-Pierre Luminet and Jeff Weeks that the universe has this topology. Furthermore, if a section of space-time is modeled as a four-dimensional dodecahedron, made up of 120 three-dimensional dodecahedra, then a circuit along one of the six axes is made up of 20 of these three-dimensional dodecahedra. If the minimal twist of 36 degrees is used to line up each of the 20 cells, then a full circuit is 720 degrees, an intriguing number for physicists.

Penrose's fundamental entity is a space made up of light rays—that is, an empty space with the built in metric of special relativity. In his essay "On the Origins of the Twister Theory," Penrose writes: "I think that I had very much come around to the view that massless particles and fields were to be regarded as more fundamental than massive ones."[9] One elegant model accommodates the splitting of light rays into past or future, accounts for their different chirality, and models the space of special relativity—what Penrose considers the starting point of all physics.[10]

[9] Roger Penrose, "On the Origins of the Twister Theory," *Gravitation and Geometry* (Naples: Bibliopolis, 1987), section 6.
[10] In Penrose's Twistor formulation there is an elusive and highly technical but still geometric reason (one that eludes me) for understanding why the future is more likely to occur next rather than the past: why

<center>***</center>

In 1964 Penrose was teaching at Birkbeck University, where the controversial and peripatetic David Bohm was professor of theoretical physics. I ask the reader to imagine a dramatic moment when a baton is passed from the older physicist to the younger as Penrose realizes that his problem with the entropic splitting of physical events to favor the future is the same as Bohm's problem with non-local causality. Two participants in that "moment" remember it differently from each other, so perhaps my imagined meeting of minds is apocryphal. In his email to me, David Peat wrote that "Penrose was in the math dept at Birkbeck and Bohm in the physics—in separate buildings. I went to Penrose's weekly lectures and asked him what he thought about Bohm. Despite being in the same college they didn't meet so I arranged for Penrose to give a lecture in Bohm's department. For the rest of the time I was there I never remember them meeting again. So I don't think Bohm was a strong influence on Penrose's thinking."

On the other hand, in an email written to me that same day, Basil Hiley remembers: "Penrose used to come to our seminars when he was in the Maths Department in the 60s and 70s and it could well be that we drew his attention to the question of non-locality. Also David Butt and Alan Wilson, together with their student, were doing pre-Aspect experiment in our Labs to test this notion at that time."

Whatever the details of the meetings between Bohm and Penrose, it does seem that it was at this time Penrose discovered that his formulation of twistors (as the basic element of physics) do have non-local properties. As he later wrote: "Another guiding

---

positive, future pointing, right-handed chirality is more likely the next instant than negative, past pointing, left-handed chirality. "This may look a little lopsided, as after all GR [General Relativity] is left-right symmetric. But this may not be such a bad thing, as Nature herself is left-right asymmetric." Hawking and Penrose, 1996, p.114.

principle behind twistor theory is quantum *non-locality*. We recall from the strange EPR effects . . . and more specifically from the role of 'quanglement' [quantum entanglement], as manifest particularly in the phenomenon of quantum teleportation . . . that physical behavior cannot be fully understood in terms of entirely local influences of the normal 'causal' character. This suggests that some theory is needed in which such non-local features are incorporated."[11]

In other words, Penrose discovered that the big bonus of thinking in terms of projective space is that it is inherently non-local and time-independent, and "idealized light rays (or their generalization, with spin) appear to be, in a sense, the carriers of quanglement.[12] If Penrose is right, then simply by virtue of projection, empty space has Minkowski metrics, spin, and non-locality.

\*\*\*

Nicolas deBrujn did not use quasicrystal geometry as a general description of space, nor did he tie his algorithm to other aspects of physics.[13] But in the context of Bohm and Penrose, it is instructive to do so. In the case of the three-dimensional quasicrystal, the algorithm starts with a pattern, a tessellation of six-dimensional hypercubes. Each six-dimensional hypercube is composed of five-dimensional hypercubes, each of which is composed of four-dimensional hypercubes, each of which, in turn, is composed of regular cubes, set in six-dimensional space. A single one of these regular, right-angle cubes is selected from this dense tessellation and then multiplied by a projection matrix to deliver one of two differently skewed cubes, oriented, and placed in three-dimensional space. With repeated applications of the algorithm, a three-dimensional quasicrystal will be made. The algorithm, then, has two procedures: selection and projection, and the order in which these procedures is employed is irrelevant to the outcome. A clever programmer could write a single program and ask the user which procedure to call first. The projection matrix is a given in the algorithm; it is the collection of rays normal to the faces of a dodecahedron. Think of these rays as the origin of our starting six-dimensional regular tessellation now bent by projection into three-dimensional space. What were 90 degree angles are now the irrational number arccosine one over the square root of five, about 63.44 degrees. The devil is in the selection procedure, which is too complicated to discuss here and anyway irrelevant to the discussion that follows.

---

[11] Roger Penrose, *The Road to Reality: A Complete Guide to the Laws of the Universe* (London: Jonathan Cape, 2004), p. 963.

[12] Hawking and Penrose 1996, p. 114.

[13] Nicolas deBrujn, "The Algebraic Theory of Penrose's Non-Periodic Tiling of the Plane," *Koninklijke Nederlands Akademie van Wetenschappen Proceeding Series* (1981).

The mystery of quasicrystal is that given the blocks (the two differently skewed cubes), there are no local matching rules that work perfectly to assemble a multi-unit quasicrystal. Following face-to-face matching rules of how to tessellate the blocks quickly leads to contradictions after surprisingly few blocks are placed. More complicated rules that define which faces may be put together and further define permissible vertex assemblies—and also permissible regional assemblies—such rules *can* make larger, but still not infinite, tessellations. These complicated matching rules are unwieldy, and it is hard to imagine how such rules could have a physical interpretation. Thus we have real-life quasicrystals that seem to self-assemble with a global method, modeling a non-local phenomenon.

However, remembering the original, pre-projected, six-dimensional cubic lattice the non-local property of quasicrystals is less mysterious. If you know one cubic cell, you know the 100th or the 1,000th; you know its size, shape, orientation, and how it connects with its neighbors. If a projection could retain this information, then the seemingly quixotic, non-local quasicrystal is logical after all.

In their writings, Bohm and Penrose want to start physics over so that it can be non-local yet causal: Bohm with a primacy on the implicate order and Penrose with a primacy on empty space with a special relativity metric. Both physicists want to connect the local with the global in a tighter way than is now understood, and in so doing secure an arrow of time pointing toward the future, retain entropy, and put causality on a stronger footing. Bohm wants to enfold all of space in each part of space. Penrose wants to collapse exploding light cones into spheres and light rays into projective points. Penrose's system is self-consciously a projection method, and Bohm's is an implicit projection method. Perhaps the baton that was passed in 1964 at Birkbeck College is the concept of a vastly more ambitious sense of space, the conviction that physical existence is in higher-dimensional space, even though our experience is lower-dimensional. And mediating between existence and experience are the metrics, mechanics, and methodology of projective geometry.

Tony Robbin, New York City, 2014